\documentstyle[12pt,epsf]{article}
\font\sf=cmss10                    

\makeatletter
\@addtoreset{equation}{section}

\makeatother

\def\del{\partial}
\def\Dslash{\not{\hbox{\kern-4pt $D$}}}
\def\dslash{\not{\hbox{\kern-2pt $\del$}}}
\def\gslash{\not{\hbox{\kern-2pt $\gamma$}}}
\def\g5{\gamma_5}

\newcommand{\mathbold}[1]{\mbox{\boldmath $\bf#1$}}
\def\mJ{\mathbold{J}}

\def\bbbz {{\sf Z\!\!\!Z}}

\def\sl2z{SL(2,\bbbz)}
\def\fracs#1#2{\textstyle\frac #1#2}
\def\gcd{{\cal GCD}}
\def\fracs#1#2{\textstyle\frac#1#2}

\newcommand{\be}{\begin{equation}}
\newcommand{\ee}{\end{equation}}
\newcommand{\bea}{\begin{eqnarray}}
\newcommand{\eea}{\end{eqnarray}}

\newcommand{\bean}{\begin{eqnarray*}}
\newcommand{\eean}{\end{eqnarray*}}
\newcommand{\myref}[1]{(\ref{#1})}

\begin{document}
\newcommand{\onefigure}[2]{\begin{figure}[htbp]
\begin{center}\leavevmode\epsfbox{#1.eps}\end{center}\caption{#2\label{#1}}
         \end{figure}}
\newcommand{\figref}[1]{Fig.~\protect\ref{#1}}
\setlength{\parskip}{1.35ex}
\setlength{\parindent}{0em}

{\flushright{\small MIT-CTP-2754\\
hep-th/9807117\\
\today \\}}

\vspace{1.0in}
\begin{center}\LARGE
{\bf Self-Intersection Number of BPS Junctions in Backgrounds of Three and Seven-Branes \\}
\end{center} \vskip 0.8cm
\begin{center}
{\large  Amer Iqbal }
\vskip 0.2cm
{{\it Center for Theoretical Physics,\\
LNS and Department of Physics,\\
Massachusetts Institute of Technology\\
Cambridge, MA 02139, U.S.A.}}
\vskip 0.3cm
{E-mail: iqbal@ctp.mit.edu}
\end{center}

\vspace{0.5in}
\begin{abstract}
In a recent paper DeWolfe et al. have shown how to use the 
self-intersection number of junctions to constrain the BPS 
spectrum of $\cal{N}$=2, $D=4$ theories with $ADE$ flavor symmetry 
arising on a single D3-brane probe in a 7-brane background. Motivated 
by the existence of more general $\cal{N}$=2, $D=4$ theories 
arising on the worldvolume of multiple D3-brane probes we 
show how to compute the self-intersection number of junctions 
in the presence of 7-branes and multiple D3-branes. 
\end{abstract}
\newpage

\section{Introduction}
String junctions \cite{aharony} play an important 
role in understanding non-perturbative aspects of Type IIB 
superstring theory. They are required  
to explain gauge symmetry enhancement when non-local 7-branes 
coincide \cite{GZ}. String junctions are constructed from 
$(p,q)$ strings \cite{witten} and are non-perturbative 
objects. 

Their role in determining the BPS spectrum of $\cal{N}$=2 theories 
in four dimensions has been studied in several 
papers \cite{ansar,amer,nekrasov}. 
The BPS states of the $\cal{N}$=2, $D$=4 SYM are realized as 
BPS string junctions connecting a D3-brane to 7-branes. In 
this probe picture the string junctions representing the BPS 
states are characterized by the invariant charges associated 
with the 7-branes \cite{DeZ}. String junctions in the presence 
of non-local 7-branes have interesting group theoretical 
properties \cite{DeZ} which can be used together with the bound 
on the self-intersection number of BPS junctions \cite{amer} to 
constrain the spectrum of $\cal{N}$=2  $D$=4 SYM with $ADE$ 
flavor symmetry. For $N_{f}\leq 4$ flavors all the states allowed 
by the constraint are realized as BPS states \cite{amer,nekrasov}. 
Three string 
junctions have also been used to construct 
 1/4 BPS states 
in the $SU(N)$ $\cal{N}$=4, $D=4$ SYM theory 
\cite{bergman,kolbergman}. The large $N$ limit of 
$\cal{N}$=2 theories arising on the worldvolume 
of $N$ D3-branes in the 7-brane background  
was studied in \cite{maldacena}. 

In this paper we will show how to calculate the 
self-intersection number of junctions in a 7-brane 
background in the presence of multiple D3-branes. Eqns \myref{intersection} 
and \myref{genus} are the main results of this paper, they can be 
used to constrain the BPS spectrum of $\cal{N}$=2, 
$D$=4 theory with $ADE$ flavor symmetry arising on 
multiple D3-branes \cite{lowe}. The results of such a 
study will be given elsewhere. 
  
A junction of Type IIB string theory compactified 
on a manifold $B\times S^{1}$ is a two dimensional 
surface in M-Theory compactified on a hyper-K\"{a}hler 
manifold $X$ which is an elliptic fibration over the 
base manifold $B$. The K\"{a}hler class of the elliptic 
fiber is related  to the radius $R$ of $S^{1}$ and 
is given by $\frac{l_{p}^{3}}{R}$, where $l_{p}$ 
is the eleven-dimensional Planck length. 
If there are no  7-branes the fibration is trivial 
and the manifold $X$ is just the product 
$B\times T^{2}$. In the presence of 7-branes the 
fibration is non-trivial and the location of 
the 7-branes on the base $B$ determines the location 
of singular fibers \cite{vafa}. Type IIB string theory on $B$ is 
obtained from M-Theory on $X$ by taking the 
K\"{a}hler class of the elliptic fiber to zero. 
 
BPS junctions on the base $B$ are made of strings 
lying along geodesics. These are obtained by wrapping 
M2-branes around holomorphic curves of $X$ 
\cite{krogh}. A $(p,q)$ string 
corresponds to an M2-brane wrapped around a 
holomorphic curve which is a product of a $(p,q)$ 
cycle of the elliptic fiber and a geodesic in the 
base representing the position of the string in 
the $R \longrightarrow \infty$ limit.

With every junction $\mJ$ of the Type IIB string 
theory one can associate a two dimensional surface 
$J$. The correspondence is such that in the limit 
$R\longrightarrow \infty$ the surface $J$ goes to 
junction $\mJ$ on the base. This correspondence 
allows us to associate with every pair of 
junctions $\mJ$ and $\mJ'$ a geometrical quantity, 
the intersection number $^{\#}(J\cdot J')$,
\be
(\mJ,\mJ') \equiv \mbox{}^{\#}(J\cdot J').
\ee

 Since the self-intersection number of a holomorphic 
curve $J\subset X$ of genus $g$ with $b$ boundary 
components is given by \cite{Hartshorne}
\be 
^{\#}(J\cdot J)=2g-2+b,
\label{maineqsec1}
\ee
BPS junctions satisfy a constraint, $(\mJ,\mJ) \geq -2 $.

The self-intersection number of a junction can be calculated in 
terms of  $(p,q)$ charges of the strings forming the 
junction therefore (1.2) allows 
us to determine the topology of the curve corresponding 
to the BPS junction in terms of the charges. 
We illustrate with a detailed argument how the 
self-intersection number of a three string junction 
is related 
to an $\sl2z$ invariant number associated with the 
junction \cite{DeZ,kolbergman}.
This $\sl2z$ 
invariant plays an important role in characterizing 
BPS junctions but lacks a direct geometrical 
interpretation in Type IIB string theory.

\section{Self-intersection number and the $\sl2z$ invariant}

In this section we will relate the invariants of the junctions 
with the intersection numbers of the corresponding curves in 
M-theory. The tension $T_{p,q}$ of a $(p,q)$ string is related 
to the length of the 1-cycle of the torus on which the M2 brane 
is wrapped. Since
\be
T_{p,q} = \frac{1}{\sqrt{\tau_{2}}}|p-q\tau|,
\ee
therefore a $(p,q)$ string corresponds to the 1-cycle $C_{p,q}= p\alpha-q\beta$. Where $\alpha$ and $\beta$ are the canonical basis 
homology cycles of the torus satisfying $^{\#}(\alpha\cdot \beta)=1$.

 With a junction $\mJ$ in IIB we associate a 2-cycle $J$ in M-theory 
such that in the limit $R \longrightarrow \infty$ $J$ goes to $\mJ$. We 
associate with two junctions $\mJ$ and $\mJ'$ an $\sl2z$ invariant number 
 $(\mJ,\mJ')$. We define this number to be the intersection number of the 
corresponding curves in M-theory \cite{nekrasov,DeZ,amer},
\be
(\mJ,\mJ') = \mbox{} ^{\#}(J\cdot J').
\ee

\onefigure{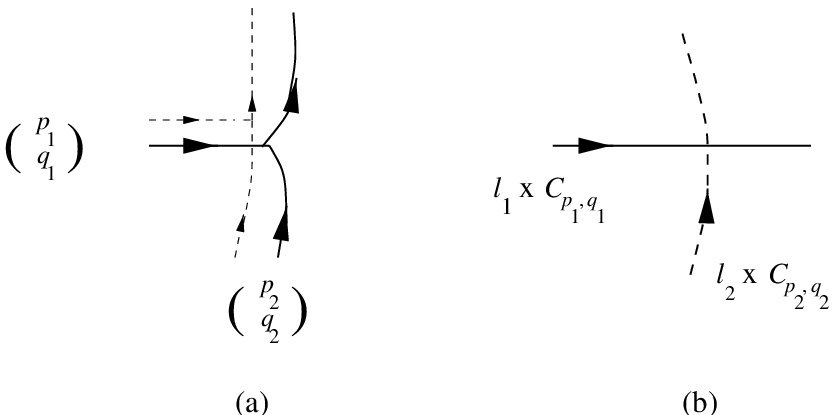}{(a) Self-intersection number is 
calculated by deforming the junction as shown. (b) Near 
the intersection point the curves have product structure.}

Consider the string junction $\mJ$  shown in \figref{junction1}(a). 
The self-intersection number gets contribution from only one 
point where a $(p_{1},q_{1})$ string crosses an $(p_{2},q_{2})$ 
string as shown in \figref{junction1}(b). Near the intersection 
point the corresponding two dimensional surfaces representing 
the $(p_{1},q_{1})$ and the $(p_{2},q_{2})$ strings in M-theory 
have product structure, $l_{i}\times C_{p_{i},q_{i}}$, where 
$l_{i}$ is the one dimensional curve on the base. 

 Then by standard manipulation \cite{griffiths&harris}
$$
(\mJ,\mJ)=\mbox{}^{\#}(J\cdot J)= \mbox{}^{\#}((l_{1}\times 
C_{p_{1},q_{1} })\cdot ( l_{2}\times C_{p_{2},q_{2}}))$$
\be
= -^{\#}(l_{1}\cdot l_{2}) ^{\#}(C_{p_{1},q_{1}}\cdot C_{p_{2},q_{2}}).
\ee

To calculate the intersection number for these 1-cycles 
we need an orientation on the base and on the fiber. The 
orientation on the base and the fiber comes from the complex 
structure of $X$. Since $X$ is a hyper-K\"{a}hler manifold 
there is a family of possible complex structures compatible 
with the metric.  However we are interested in the complex 
structures $I_{+}$ and $I_{-}$ in which the fiber is 
holomorphic. The base is holomorphic in the complex 
structure $I_{+}$ and $-I_{-}$.
Since the fiber is holomorphic therefore 
$^{\#}(\alpha\cdot \beta) =1$ and 
\be
^{\#}(C_{(p_{1},q_{1})}\cdot C_{(p_{2},q_{2})}) 
= -\left |\begin{array}{cc}
                      p_{1} & p_{2} \\
                      q_{1} & q_{2} \end{array}\right|.
\ee

The intersection number $^{\#}(l_{1}\cdot l_{2})$ 
depends on the orientation of the base therefore
\be
^{\#}(l_{1}\cdot l_{2}) = \pm 1,  \mbox{ in the 
complex structure $I_{\pm}$.}
\ee
Thus back in (2.4)
\be
(\mJ,\mJ) =\pm \left |\begin{array}{cc}
                      p_{1} & p_{2} \\
                      q_{1} & q_{2} \end{array}\right|,  
\mbox{ in the complex structure $I_{\pm}$.} 
\ee

Thus we see that if $J$ is holomorphic in $I_{+}$, 
$(p_{1}q_{2}-q_{1}p_{2})>0$.
In this case a BPS $(p,q)$ string is oriented in the 
direction $p-q\bar{\tau}$.
If $J$ is holomorphic in the complex structure $I_{-}$ 
then $(p_{1}q_{2}-q_{1}p_{2})<0$. In this case BPS $(p,q)$ 
string is oriented in the direction $p-q\tau$. 

 If there are 7-branes  on B the monodromy $K_{p,q}$ 
around a $[p,q]$ 7-brane is the same as the monodromy 
around an elliptic fiber whose $p\alpha-q\beta$ cycle 
is pinched. If $\tau$ is a holomorphic function of $z$ 
(the holomorphic coordinate on the base in the complex 
structure $I_{+}$) then going counterclockwise around 
the $[p,q]$ 7-brane $\tau(z)\longrightarrow K_{p,q}\tau(z)$. 
If $\tau$ had been function of $\bar{z}$ then it would have 
been transformed by $K^{-1}_{p,q}$. Thus the base is 
holomorphic in the complex structure $I_{+}$ when 
there are 7-branes on it.  

 In summary, {\it if $\mJ$ is a BPS string junction 
then the $\sl2z$ invariant associated with each 
junction point is of same sign. If it is positive 
(negative) for every junction point then the curve 
corresponding to $\mJ$ is holomorphic in the complex 
structure $I_{+}(-I_{-})$}.  Thus if in a string web 
the invariants associated with two junction points 
are of opposite sign then that string web cannot be 
BPS. {\it In the presence of 7-branes therefore the 
invariant associated with each junction point 
must be positive for a BPS string junction}.

\section{BPS junctions on 3-branes}
Consider IIB on $B\times S^{1}$ with only D3-branes 
present. Since 
there are no 7-branes there are no singular fibers 
and  $X=B\times T^{2}$ is the compactification 
manifold of M-theory.     
Consider $n$ D3-branes on $B$ and a BPS string 
junction $\mJ$ between them
such that the strings are either all incoming or 
outgoing so that $\sum_{i=1}^{n}p_{i}=\sum_{i=1}^{n}q_{i}=0$. 
We will associate with this junction a polygon  $P(\mJ)$ \cite{vafa}
in the lattice $\Gamma$ defined as
\be
\Gamma = \{(p,q)=p-q\bar{\tau} | p, q\in \bbbz\}~.
\ee
This polygon $P(\mJ)$ has vertices $\{(\sum_{i=1}^{k}p_{i},
\sum_{i=1}^{k}q_{i})|k=1,..,n\}$.
Since the self-intersection number does not get any 
contributions from the boundaries we can send the 
D3-branes to infinity and consider the resulting 
infinite string junction $\mJ'$ \cite{sennetwork}. 
Therefore the polygon $P(\mJ')$ is the same as 
$P(\mJ)$ and $ ^{\#}(J'\cdot J') = \mbox{} ^{\#}(J\cdot J) $.

From toric geometry we know the relationship 
between the area, $A(P(\mJ))$, of the polygon 
$P(\mJ)$ and the 
self-intersection number of holomorphic curve $J'$ \cite{fulton}:
\be
      ^{\#}(J'\cdot J') = 2 A(P(\mJ))=2g-2+b~,
\label{myfavourite}
\ee
where a cell of the lattice $\Gamma$ is declared 
to have unit area.
We define 
\be
d_{i}=\gcd(p_{i},q_{i})~,
\ee
where $\gcd(p,q)\geq 1$ is the greatest common 
divisor of $p_{i}$ and $q_{i}$.
The number of boundary components $b$ is given 
by \cite{amer}
\be
b= \sum_{i=1}^{n}d_{i}=l(P(\mJ))~,
\ee
where $l(P(\mJ))$ is the number of lattice points 
on the perimeter of $P(\mJ)$. The number of 
integral points $\#(P(\mJ))$ inside the polygon $P(\mJ)$
 is given by Pick's formula \cite{fulton}
 
\be
\#(P(\mJ)) = A(P(\mJ)) -\fracs{1}{2} l(P(\mJ)) +1~.
\label{eqfulton}
\ee

Thus from \myref{myfavourite} and \myref{eqfulton} 
\be
g= \fracs{1}{2}^{\#}(J\cdot J)-\fracs{1}{2}b+1=
A(P(\mJ))-\fracs{1}{2}l(P(\mJ))+1 =\#(P(\mJ))~.
\ee
\smallskip
In terms of the string charges $(p_{i},q_{i})$ we get

\be
^\#(J\cdot J)=2A(P(\mJ))=\left |\sum_{1\leq i<j\leq n}
\left |\begin{array}{cc}
                      p_{i} & p_{j} \\
                      q_{i} & q_{j} \end{array}\right | \right |
\label{maineqsec3}
\ee
 
\be
g=\#(P(\mJ)) = \mbox{$\fracs{1}{2}$}^{\#}(J\cdot J)-
\mbox{$\fracs{1}{2}$}\sum_{i=1}^{n}d_{i}+1~. 
\label{maineq2sec3}
\ee

If the junction $\mJ$ is BPS and all the strings 
are oriented in the direction $p-q\bar{\tau}$ then each term in the sum 
(2.15) is positive and the area of $P(\mJ)$ is the sum of these terms. If 
the strings are oriented in the direction $p-q\tau$ then each term in the 
sum is negative and the area is minus the sum of these negative terms. Thus 
we need to take the absolute value of the sum in \myref{maineqsec3}.  

\section{BPS junctions on 7-branes}
$[p,q]$ 7-branes of type IIB are the singular 
fibers of a hyper-K\"{a}hler  manifold 
$X$ in the lift to M-theory. 2-cycles with 
support on the 7-branes 
do not have boundaries. Therefore in the presence of 
7-branes (2.11) fails and the number of interior points 
in the polygon $P(\mJ)$ is not equal 
to the 
genus of the corresponding holomorphic curve $J$. The 
self-intersection number of a genus $g$ holomorphic curve 
$J\subset X$ 
 without boundary is [\myref{maineqsec1}]

\be
^{\#}(J\cdot J) = 2g-2~.
\ee

In fact in a homology class $J\in H^{2}(X)$  of 
self-intersection number 
greater than or equal to minus two the minimal 
area surface is a holomorphic 
curve of genus $\frac{^{\#}(J\cdot J) +2}{2}$ \cite{wolfson}. 
This implies that a 
junction corresponding to a surface of self-intersection 
number greater than or equal to minus two will 
always have a BPS representative. This BPS representative 
corresponds the curve of least genus and minimal area in 
this homology class.  It will be unique if the 
self-intersection number is minus two \cite{wolfson}. This has 
been explained in the Type IIB viewpoint in \cite{Hauer}.

Consider a junction $\mJ$ which ends on 7-branes. The 
strings ending on the 7-branes cannot be  deformed away 
from the 7-branes and therefore also contribute to the 
self-intersection number. This additional contribution 
changes the relation between the number of interior 
points in $P(\mJ)$ and the genus of $J$. If a $(p_{i},q_{i})$ 
string ends on a $[p_{i},q_{i}]$ 7-brane  then the 
contribution to the self-intersection due to the 
presence of the 7-brane is $-d_{i}^{2}$. (A $(p_{i},q_{i})$ 
string, with $p_{i}$ and $q_{i}$ relatively prime, 
ending on  a 7-brane is like a half sphere with a 
marked point. A general $(p_{i},q_{i})$ string is a 
sphere wrapped  $d_{i}$ times around itself. The 
contribution to the self-intersection number from 
one of the fixed points of the sphere is $\frac{1}{2}(-2)$. 
With $n$-prongs we get a contribution equal to one half the 
self-intersection of a sphere wrapped $n$-times around it 
self which is equal to $-n^{2}$ = $-d_{i}^{2}$.)

Since the self-intersection number does not change under a 
continuous deformation of $J$, we deform the 
junction $\mJ$ such that the strings  do not cross the 
branch cuts associated with the 7-branes.
The total self-intersection number is the sum of 
self-intersection numbers from the string endpoints 
on the 7-branes and the junction points. The  
contribution to the total self-intersection number 
from the junction points is equal to the self-intersection 
number of an infinite junction with same charges. Thus 
\be
^{\#}(J\cdot J) =2A(P(\mJ))-\sum_{i=1}^{n}d_{i}^{2}~.
\ee
The last term is the contribution from the end points 
of the strings on the 7-branes. In terms of string 
charges $(p_{i},q_{i})$ we get
\be
^{\#}(J\cdot J) =\sum_{1\leq i<j\leq n}\left |\begin{array}{cc}
                      p_{i} & p_{j} \\
                      q_{i} & q_{j} \end{array}\right|-
\sum_{i=1}^{n}d_{i}^{2}~.
\ee
This is in agreement with \cite{nekrasov,DeZ}. Using 
(2.12) we get 

\be g= \#(P(\mJ)) +\mbox{$\fracs{1}{2}$}\sum_{i=1}^{n}\{d_{i}-
d_{i}^{2}\}=\mbox{$\fracs{1}{2}$}^{\#}(J\cdot J)+1~.
\ee
 Thus we see that holomorphic curves which correspond 
to junctions with ends on the 7-branes with non-relatively 
prime charges have lower genus than a junction with same 
charges ending on D3-branes (see \myref{maineq2sec3}).

\section{BPS junctions on 3-branes and 7-branes}  

Consider a BPS junction $\mJ$ in the presence of $n$ 
7-branes and $m$ D3-branes. The corresponding 
holomorphic curve $J$ has $b$ boundary components 
(from strings ending on D3-branes) and $n$  marked 
points (from strings ending on 7-branes).  

We choose  a presentation $\mJ'$ of the junction 
$\mJ$ in which strings do not cross the branch cuts 
associated with the 7-branes and the D3-branes are 
grouped together far from the 7-branes. This is 
achieved by continuous deformation of the junction 
which leaves the self-intersection numbers invariant.
Let $\{(p_{i},q_{i}),i=1,..,n\}$ and $\{(p_{i},q_{i}),
i=n+1,..,n+m\}$ be the charges of the strings forming 
the junction $\mJ'$ and 
ending on 7-branes and D3-branes respectively. We 
define $P(\mJ')$ as the polygon with vertices 
$\{(\sum_{i=1}^{k}p_{i},\sum_{i=1}^{k}q_{i})|k=1,..,n+m\}$. 
Then
\be
^{\#}(J\cdot J) =^{\#}(J'\cdot J')= 2g-2+b=2A(P(\mJ')) 
-\sum_{i=1}^{n} d_{i} ^{2}~.
\ee
The last term is the contribution from the strings 
ending on 7-branes.
The number of boundary components $b$ is given by
\be
b=\sum_{i=n+1}^{n+m}d_{i}~, \mbox{~~~and~~~} l(P(\mJ'))-b=\sum_{i=1}^{n}d_{i}~.
\ee
In terms of string charges $(p_{i},q_{i})$ we get  
\be
^{\#}(J\cdot J) =\sum_{1\leq i<j\leq n+m}\left |\begin{array}{cc}
                      p_{i} & p_{j} \\
                      q_{i} & q_{j} \end{array}\right|-
\sum_{i=1}^{n}d_{i}^{2}~,
\label{intersection}
\ee

\be
g=\#(P(\mJ')) +\mbox{$\fracs{1}{2}$} \sum_{i=1}^{n} \{d_{i}-d_{i}^{2}\} =
\mbox{$\fracs{1}{2}$}^{\#}(J\cdot J)-\mbox{$\fracs{1}{2}$}   \sum_{i=n+1}^{n+m}d_{i}+1~.
\label{genus}
\ee

From the expression for the genus in terms of the charges 
we  see that the presence of 7-branes lift some of the moduli as 
anticipated in \cite{nekrasov}. In case of non BPS 
junction $\mJ$ \myref{genus} gives a lower bound on the genus 
of $J$. Equations \myref{intersection} and \myref{genus} are the main 
results of this paper. The utility of \myref{intersection} and \myref{genus} is that only the string prongs at the branes are needed and the explicit realization of the junction is not required.

As  an example consider a junction $\mJ_{0}$ connecting 
two D3-branes with an $SO(8)$ singularity as shown in 
\figref{junction2}. 

\onefigure{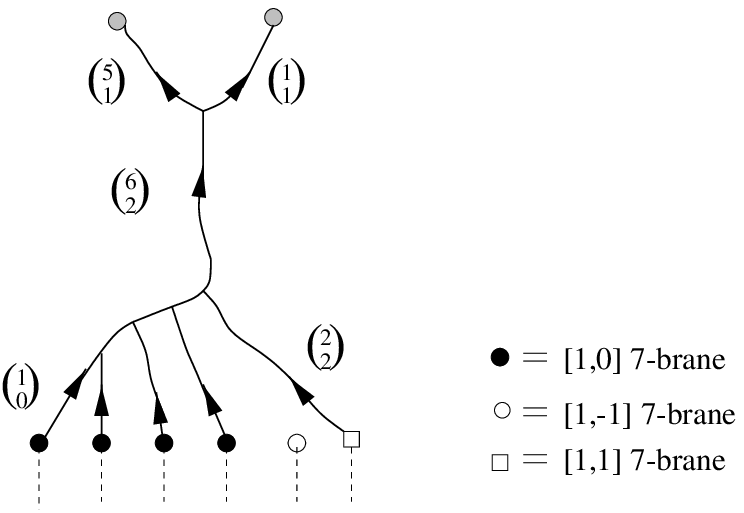}{A non BPS junction in the $\cal{N}$=2, 
$D$=4 $Sp(4)$ theory with $SO(8)$ flavor symmetry.}

BPS junctions satisfying $(\mJ,\mJ) = 2g-2+b\geq -1$
represent BPS states in the $\cal{N}$=2, $D$=4 
$Sp(4)$ theory with four flavors. This theory has been studied in \cite{leeYi}. Calculating 
the self-intersection number of $\mJ_{0}$ by adding 
contributions from  the 7-branes and the junction points we get
\be
(\mJ_{0},\mJ_{0}) = 4(-1) -2^{2}+\left |\begin{array}{cc}
                      4     & 2     \\
                      0     & 2     \end{array}\right|+\left 
|\begin{array}{cc}
                      1     & 5     \\
                      1     & 1     \end{array}\right|=-4~.
\ee

 We can use \myref{intersection} directly to calculate the self-intersection 
number. Here we have $n=5$, $m=2$, 
$\{(p_{i},q_{i})|i=1,..,7\}$ = 
$\{(1,0), (1,0), (1,0),$
$(1,0), (2,2), (-1,-1), (-5,-1)\}$ 
and $\{d_{i}|i=1,..,7\} = \{1, 1, 1, 1, 2, 1, 1\}$. One finds, of course the same result  $(\mJ_{0},\mJ_{0}) =-4$.
Therefore $\mJ_{0}$ is not a BPS junction.

\subsection*{Acknowledgments}

I  would like to thank  Oren Bergman, Oliver DeWolfe, 
Tam\'as Hauer and Barton Zwiebach for useful discussions. 
This work was 
supported by the U.S.\ Department of Energy under contract
\#DE-FC02-94ER40818.

\end{document}